\documentclass[twocolumn,english,reprint,amsmath,amssymb,aps]{revtex4-1}
\usepackage[T1]{fontenc}
\usepackage[latin9]{inputenc}
\usepackage{verbatim}
\usepackage{esint}
\usepackage{color}

\makeatletter
\usepackage{graphicx}% Include figure files
\usepackage{dcolumn}% Align table columns on decimal point
\usepackage{bm}% bold math
\makeatother

\usepackage{babel}
\begin{document}
\preprint{APS/123-QED}

\title{Derive Lovelock Gravity from String Theory in Cosmological Background}

\author{Peng Wang}
\email{pengw@scu.edu.cn}
\author{Houwen Wu}
\email{iverwu@scu.edu.cn}
\author{Haitang Yang}
\email{hyanga@scu.edu.cn}
\affiliation{College of Physics, Sichuan University, Chengdu, 610065, China}

\author{Shuxuan Ying}
\email{ysxuan@cqu.edu.cn}
\affiliation{Department of Physics, Chongqing University, Chongqing, 401331, China}

\date{\today}

\begin{abstract}

\noindent It was proved more than three decades ago, that the first order $\alpha'$ correction of string effective theory could be written as the Gauss-Bonnet term, which is the quadratic term of Lovelock gravity. In cosmological background, with an appropriate field redefinition, we reorganize the infinite $\alpha'$ corrections of string effective action into a finite term expression for any specific dimension. This finite term expression matches Lovelock gravity exactly and thus fix the couplings of Lovelock gravity by the coefficients of string effective action. This result thus provides a strong support to string theory.

\end{abstract}

\maketitle

Physically, an extension of general relativity requires the generalized Einstein
tensor to respect three conditions:  1) It is symmetric;
2) It is  divergence free (Bianchi identity); 3) Higher derivatives  of the metric
are absent in the EOM (ghost free). Lovelock gravity \cite{Lovelock:1971yv} is the most general metric theory extension
satisfying these conditions.
In $D=d+1$ dimensional spacetime,
it is constructed by  dimensionally extended Euler densities (setting $16\pi G_{D}=c=1$):

\begin{eqnarray}
I_{Love} & = & \int d^{D}x\sqrt{-g}\sum_{k=0}^{\left[\frac{D-1}{2}\right]}\alpha_{k}\lambda^{2k-2}\mathcal{L}_{k},\nonumber \\
\mathcal{L}_{k} & \equiv & \frac{1}{2^{k}}\delta_{\rho_{1}\cdots\rho_{k}\sigma_{1}\cdots\sigma_{k}}^{\mu_{1}\cdots\mu_{k}\nu_{1}\cdots\nu_{k}} R_{\mu_{1}\nu_{1}}^{\quad\;\rho_{1}\sigma_{1}}\cdots R_{\mu_{k}\nu_{k}}^{\quad\;\rho_{k}\sigma_{k}},\label{eq:Love action}
\end{eqnarray}

\noindent where $\left[\left(D-1\right)/2\right]$ denotes the integer
part of $\left(D-1\right)/2$, $\delta_{\rho_{1}\cdots\rho_{k}\sigma_{1}\cdots\sigma_{k}}^{\mu_{1}\cdots\mu_{k}\nu_{1}\cdots\nu_{k}}$
is the generalized anti-symmetric  Kronecker delta, $\alpha_{k}$ are
real dimensionless parameters and we have isolated a length (Planck) scale $\lambda$ for later convenience.
This action (\ref{eq:Love action}) has a finite number of terms for any finite spacetime dimension. Particularly, it reduces to the Einstein action at $D=4$ with $\alpha_{0}\lambda^{-2}=-2\Lambda$
and $\alpha_{1}=1$. The next term is the Gauss-Bonnet gravity.

On the other hand, the (closed) string effective action receives higher derivative corrections, controlled by the squared string length $\alpha'$. These higher derivative corrections lead to ghost degrees of freedom. Since string theory itself is ghost free, it is believed that the higher derivative pathology is caused by artificial truncation and can be cured by arranging the terms properly. Therefore, Lovelock gravity should somehow can be derived from the string effective action, if string theory is the right candidate of full quantum gravity theory.

In  \cite{Zwiebach:1985uq}, the quadratic Gauss-Bonnet term of Lovelock gravity was indeed reproduced from the first order $\alpha^{\prime}$ correction of the string effective action by field redefinitions. However,  the traditional approaches seem hopeless to eliminate the higher derivatives in  higher order $\alpha'$ corrections by field redefinitions. Thus deriving Lovelock gravity from string effective action seemed not possible until recently.

Thanks to the recent remarkable progress on classifying all the
$\alpha^{\prime}$ corrections done by Hohm and Zwiebach \cite{Hohm:2015doa,Hohm:2019ccp,Hohm:2019jgu},
it becomes possible to study whether Lovelock gravity matches string theory.
In 1990s, Meissner and Veneziano showed
that, to the first order in $\alpha^{\prime}$, when all fields only
depend on time, the   string effective action has
an explicit $O\left(d,d\right)$ symmetry \cite{Veneziano:1991ek, Meissner:1996sa}.
Sen proved this is true to all orders in $\alpha^{\prime}$ and for
configurations independent of $m$ coordinates, the symmetry is $O\left(m,m\right)$
\cite{Sen:1991zi,Sen:1991cn}. This is also confirmed in ref. \cite{Meissner:1991zj}
from the perspective of $\sigma$ model expansion. It turns out that
to the first order in $\alpha^{\prime}$, the $O\left(d,d\right)$
matrix can maintain the standard form in terms of $\alpha^{\prime}$
corrected fields, for time dependent configuration \cite{Meissner:1996sa}.
One can be easily convinced that this is also true for all orders
in $\alpha^{\prime}$, from the derivations in \cite{Meissner:1996sa}.
Based on this assumption, Hohm and Zwiebach \cite{Hohm:2015doa,Hohm:2019ccp,Hohm:2019jgu}
showed that, for cosmological configurations, the $\alpha^{\prime}$
corrections to all orders, can be put into incredibly simple patterns.
The dilaton  appears trivially and only first order
time derivatives need to be included. In other words, \emph{higher
derivative} corrections are absorbed into redefined fields by $O\left(d,d\right)$
invariant field redefinitions. In our recent work \cite{Wang:2019mwi},
we found that the classification of $\alpha^{\prime}$ corrections also
applies to   space-dependent background (domain-wall
ansatz). We also presented   non-singular string cosmological solutions  in refs. \cite{Wang:2019kez,Wang:2019dcj}.

The purpose of this letter is to derive Lovelock gravity from  string effective action in cosmological background. To this end, we start with  the FLRW metric:

\begin{equation}
ds^{2}=-dt^{2}+ a\left(t\right)^{2}dx_{i}dx^{i}.\label{eq:FLRW in Love}
\end{equation}

\noindent The Lovelock action (\ref{eq:Love action}) is simplified as

\begin{eqnarray}
I_{Love} & = & \int dt\, a^{D-1}\sum_{k=0}^{\left[\frac{D-1}{2}\right]}\left(-\frac{1}{2k-1}\right)\alpha_{k}\lambda^{2k-2}\times\nonumber \\
 &  & \frac{\left(D-1\right)!}{\left(D-2k-1\right)!} H^{2k}\nonumber  \\
 &=& \int dt\, a^{D-1}\Big[\alpha_0 \lambda^{-2} -\alpha_1 (D-1)(D-2) H^2 \nonumber\\
 & &-\frac{\alpha_2\lambda^2}{3} (D-1)(D-2)(D-3)(D-4) H^4 +\cdots \Big],\nonumber \\
\label{eq:simp love action}
\end{eqnarray}

\noindent where we modulo the spatial volume factor, and $H\equiv\dot a\left(t\right)/ a\left(t\right)$
is the Hubble parameter. The solutions and relevant discussions
of this action can be found in ref. \cite{Deruelle:1989fj}.

We then turn to the  string effective action
corrected by all orders in $\alpha^{\prime}$:

\begin{eqnarray}
I_{string} & = & \int d^{D}x\sqrt{-\tilde g}e^{-2  \phi}\left(\tilde R+4\left(\partial  \phi\right)^{2} \right.\nonumber \\
 &  & \left.+ \frac{1}{4}\alpha^{\prime}\left(\tilde R^{\mu\nu\rho\sigma} \tilde  R_{\mu\nu\rho\sigma}+\ldots\right)+\alpha^{\prime2} \left(\ldots\right)+\ldots\right),\nonumber \\
\label{eq:ordinary action}
\end{eqnarray}

\noindent where we use ``tilde'' to indicate the string frame with the string metric  $\tilde g_{\mu\nu}$, $\phi$ is the
physical dilaton and we set Kalb-Ramond field $b_{\mu\nu}=0$ for
simplicity.

Hohm and Zwiebach showed that  \cite{Hohm:2019ccp,Hohm:2019jgu}, in the FLRW ansatz, this action could be dramatically
simplified as

\begin{equation}
I_{HZ}=\int dte^{-\Phi}\left(-\dot{\Phi}^{2}-d\sum_{k=1}^{\infty}\left(-\alpha^{\prime}\right)^{k-1}2^{2k+1}c_{k} \tilde H^{2k}\right),\label{eq:HZ action}
\end{equation}
where $\Phi\equiv 2\phi-\ln\sqrt{\det \tilde g_{ij}}$ is the  $O\left(d,d\right)$ invariant dilaton. For bosonic string theory, we have $c_{1}=-\frac{1}{8}$ exactly matching Einstein gravity and $c_{2}=\frac{1}{64}$ fixing the coefficient of the Gauss-Bonnet term. $c_{k\geq3}$ are yet unknown constants.
It is worth noting that the action (\ref{eq:HZ action}) has an explicit scale-factor duality

\begin{equation}
\tilde a\left(t\right)\rightarrow \tilde a^{-1}\left(t\right),\tilde H\left(t\right)\rightarrow-\tilde H\left(t\right),\Phi\left(t\right)\rightarrow\Phi\left(t\right),\label{eq:ODD}
\end{equation}
which belongs to the more general   $O\left(d,d\right)$ transformations.

To compare with Lovelock gravity, we need to transform  the Hohm-Zwiebach action (\ref{eq:HZ action})
to  Einstein frame by
\begin{equation}
\tilde g_{\mu\nu}=e^{-2\omega}{g}_{\mu\nu},\quad\omega\equiv-\frac{2\left(\phi-\phi_{0}\right)}{D-2},\label{eq:trans 1}
\end{equation}
where  $\phi_{0}$ is the expectation of the dilaton. However, since Lovelock gravity (\ref{eq:simp love action}) is a pure metric theory without dilaton degree of freedom, here the string frame is same as Einstein frame. So one can simply set  $\phi=\phi_0$ in the Hohm-Zwiebach action (\ref{eq:HZ action}) to obtain

\begin{eqnarray}
{I}_{HZ} & = & e^{-2\phi_{0}}\int dt {a}^{D-1}\left[-\left(D-1\right)\left(D-2\right) {H}^{2}\right.\nonumber \\
 &  & \left.-\left(D-1\right)\sum_{k=2}^{\infty}\left(-\alpha^{\prime}\right)^{k-1}2^{2k+1}c_{k} {H}^{2k}\right].\label{eq:HZ action final}
\end{eqnarray}
It is worth noting that the $O(d,d)$ symmetry (\ref{eq:ODD})  no longer exists after the dilaton being fixed.

At first glance, it seems this string effective action  is completely different from Lovelock gravity  (\ref{eq:simp love action}), since for a specific dimension $D$, Lovelock gravity has a finite number of $[\frac{D+1}{2}]$  terms, but the string effective action has infinitely many terms for any dimension.

The point is to note that for a field theory, we are always free to make whatever regular field redefinitions which do not alter  the   $S$-matrix. So, if two theories can be connected by some regular field redefinitions, they are identical. For  the string effective action, even at the order of ${\alpha'}^2$ term, there  already exist ambiguities of field redefinitions. These ambiguities  simply reflect the freedom to arrange higher orders in  ${\alpha'}$. Therefore, to prove the equivalence of  Lovelock gravity  and string theory, we only need to find some regular field redefinition which transforms eq. (\ref{eq:HZ action final}) to the same pattern of eq. (\ref{eq:simp love action}). This can be done order by order in $\alpha'$ as follows. Substituting

\begin{equation}
{H}^{2} \to {H'}^{2}={H}^{2}+\sum_{n=2}^{\infty}\left(-\alpha^{\prime}\right)^{n-1}A_{n}{H}^{2n},
\end{equation}
into  eq. (\ref{eq:HZ action final}),  with the coefficients $A_n$,
\begin{eqnarray}
A_{2} & = & -2^{5}c_{2}\left(\frac{1}{\left(D-2\right)}+\frac{1}{6}\left(D-4\right)\left(D-3\right)\right),\nonumber \\
A_{3} & = & -2^{7}c_{3}\left(\frac{1}{\left(D-2\right)}\right.\nonumber \\
 &  & \left.-\frac{1}{24}\left(D-6\right)\left(D-5\right)\left(D-4\right)\left(D-3\right)\right)\nonumber \\
 &  & +2^{11}c_{2}^{2}\left(\frac{1}{\left(D-2\right)^{2}}+\frac{1}{6}\frac{\left(D-4\right)\left(D-3\right)}{\left(D-2\right)}\right),\nonumber \\
&\cdots&
\label{eq:Per F Def}
\end{eqnarray}
one can easily verify that eq. (\ref{eq:HZ action final}) fits into the same pattern of  Lovelock gravity eq. (\ref{eq:simp love action}). Then we are able to identify the relation between $c_k$ and $\alpha_k$, which will be given in eq. (\ref{eq:result}).

Note there is no constant  (cosmological)  term in critical string theory, so our matching starts from $H^2$ term. The factor $1/(D-2)$ in  denominators causes no trouble since nontrivial gravity exists in $D>2$.

Using a trick introduced by Hohm-Zwiebach in  \cite{Hohm:2019ccp,Hohm:2019jgu}, we can figure out the non-perturbative form of the above field redefinition by organizing the terms   appropriately. We first set

\begin{eqnarray}
\delta_{k}{I}_{HZ} & = &  \left(-\alpha^{\prime}\right)^{k-1}e^{-2\phi_{0}}\int dt  {a}^{D-1}2^{2k+1}c_{k}\left(D-1\right)\times\nonumber \\
 &  & \left(\left(-\right)^{k}\frac{1}{\left(k+1\right)!}\frac{\left(D-2\right)!}{\left(D-2k-1\right)!} + 1\right) {H}^{2k},\nonumber \\
\label{eq:k field redefinitions}
\end{eqnarray}

\noindent where

\begin{equation}
\frac{\left(D-2\right)!}{\left(D-2k-1\right)!}=\left(D-2k\right)\ldots\left(D-3\right)\left(D-2\right),
\end{equation}
is naturally truncated at $k=[(D-1)/2]$. Note the factor $\left(-\right)^{k}\frac{1}{\left(k+1\right)!}$
is used to match the coefficients of Hohm-Zwiebach action at each
order. Then it is not hard to verify that the non-perturbative expression of the field redefinition (\ref{eq:Per F Def}) is

\begin{eqnarray}
I_{HZ}&\to& {I}_{HZ}^{\prime}  =  {I}_{HZ}+\sum_{k=2}^{\infty}\delta_{k} {I}_{HZ} \nonumber \\
 & = & \int dt {a}^{D-1}\sum_{k=1}^{\left[\frac{D-1}{2}\right]}\beta_{k}\left(\sqrt{\alpha^{\prime}}\right)^{2k-2}\frac{\left(D-1\right)!} {\left(D-2k-1\right)!} {H}^{2k},\nonumber \\
\label{eq:simp HZ action}
\end{eqnarray}

\noindent where
\begin{equation}
\beta_{k}=-e^{-2\phi_{0}}\frac{1}{\left(k+1\right)!}2^{2k+1}c_{k}.
\end{equation}
We redefined $c_{1} =\frac{1}{4}$ to replace its original value $-\frac{1}{8}$  to make the expressions simpler. Comparing with Lovelock gravity (\ref{eq:simp love action}), it is ready to find

\begin{equation}
\alpha_{k}=\frac{2k-1}{\left(k+1\right)!}2^{2k+1}c_{k},\label{eq:result}
\end{equation}
after identifying $\lambda \equiv \sqrt{\alpha'}$. The coupling constants $\alpha_k$ in Lovelock gravity were arbitrary. However, we now see that as an effective theory of string theory, these couplings are fixed by the corresponding coefficients in the string effective action.

One enlightening observation from our derivation is that,   when a proper field redefinition is used, infinitely many $\alpha'$ corrections are replaced by finite terms for a specific dimension, say $D=26$. If this is also true when the dilaton is turned on, we are able actually to get a closed expression for the string  spacetime action. Therefore, we may fix the non-perturbative string theory through perturbative results.

Moreover, we wish to address a closely related subject, the quasi-topological gravity, which was first proposed
by  Myers and  Robinson \cite{Myers:2010ru} under a specialized black hole ansatz.  In ref.
\cite{Dehghani:2013oba,Cisterna:2018tgx,Arciniega:2018fxj},
the quasi-topological gravity was extended to FLRW background. As we have mentioned,
Lovelock theorem asserts that there is no alternative generalization
of Einstein's gravity, which still possesses the second-order equations
of motion. Myers and  Robinson adopted a different way. They first
select all possible curvature-cubed interactions.  Then, all these
curvature-cubed interactions are combined together with undetermined coefficients.
For some special metric ansatz, it is possible to fix the relations between
coefficients to guarantee the equations of motion to be second-order.
Finally, using the interactions with fixed coefficients, the ghost-free
generalised action for that special metric ansatz was obtained. Therefore, the
quasi-topological gravity could allow   cubic terms  active
in five dimensions, but only for some preselected special background, as opposed to cubic
Lovelock gravity   only exists in seven dimensions for general
background. In addition to
quasi-topological gravity, there are some other higher-curvature generalizations
of gravity \cite{Arciniega:2018tnn,Cano:2020oaa}.

Based on our derivation, it is possible to ask whether
these different kind of generalizations of gravity could be related
to Hohm-Zwiebach action by appropriate field redefinitions.
The answer is tricky. It should be noted there are two sorts of   coefficients
in string effective action, namely,  ambiguous and unambiguous coefficients.
The unambiguous coefficients   are independent
of   field redefinitions, while the ambiguous coefficients are altered by
field redefinitions. The observables of string theory, say S-matrix,
are controlled by unambiguous coefficients. This is why the field
redefinitions  do
not affect physical results. In Hohm and Zwiebach's work,
a series of field redefinitions
were used to set all ambiguous coefficients vanishing altogether, and only the
unambiguous coefficients
$2^{2k+1}c_{k}$ are left to   dramatically simplify the action. To keep the coefficients
$2^{2k+1}c_{k}$ invariant, the $O\left(d,d\right)$ invariant field redefinitions, namely $\delta_{k}I_{HZ}$,
only can provide following terms to modify   ambiguous coefficients
in the low energy effective action:

\begin{equation}
\left[\sum_{i=0}^{2k-1}a_{i}D^{i}\right]\left(\sqrt{\alpha^{\prime}}\right)^{2k-2}H^{2k},\label{eq:FD}
\end{equation}

\noindent where $a_{i}$ are arbitrary constants and $D^{i}$ denotes
an $i$-th power of the spacetime dimension $D$. It is worth noting
that $a_{i}$ do not depend on the spacetime dimension. However,
a curvature-cubed term of the quasi-topological gravity
takes a form:

\begin{equation}
\left[\sum_{i=0}^{5}\frac{a_{i}}{\left(2D-3\right)}D^{i}\right]\mu_{3}H^{6},\label{eq:quasi cubed term}
\end{equation}

\noindent where $\mu_{3}$ is an arbitrary dimensional parameter with dimension  $\left(\mathrm{Length}\right)^{4}$, and
it does not depend on the spacetime dimension. We can see that the unambiguous coefficients are absent in (\ref{eq:quasi cubed term})  and
there is no way to find an appropriate
$O\left(d,d\right)$ invariant field redefinition
to cancel the $2D-3$ denominator in (\ref{eq:quasi cubed term}) to match eq. (\ref{eq:FD}). Therefore, we
cannot reach the quasi-topological gravity   by field redefinitions from Hohm-Zwiebach
action.

Moreover, we need to note that  generalizing Hohm-Zwiebach
action to a general background is not available yet,  because the $O\left(d,d\right)$
symmetry constraints the ansatz of background. We believe this problem
could be solved by double field theory.

In summary, we showed that Lovelock gravity can be derived from Hohm-Zwiebach action via an appropriate field redefinition. The couplings of Lovelock gravity are fixed by the coefficients of string effective action.  Although our discussion only applies to cosmological background, it does provide a very strong evidence for string theory. Generalizing our discussion to situations that rely on two or more coordinates are very desirable. To this end, we  need to extend Hohm-Zwiebach action to depending on more than just one variable. The first nontrivial step on this track recently has been done in
\cite{Eloy:2019hnl,Eloy:2020dko}.

\begin{acknowledgments}
This work is supported in part by the NSFC (Grant No. 11875196, 11375121, 11005016 and 11947225).

\end{acknowledgments}

\end{document}